\documentclass[12pt]{article}
\usepackage{eurosym}
\usepackage{natbib}
\usepackage{setspace}
\usepackage{amsmath}
\usepackage{graphicx}
\usepackage{float}
\usepackage{hyperref}
\usepackage{xcolor}
\usepackage[top=2cm,bottom=4cm,left=3cm,right=3cm,asymmetric]{geometry}
\usepackage{amssymb}
\usepackage[labelfont=bf]{caption}

\newtheorem{proposition}{Proposition}
\newtheorem{assumption}{Assumption}

\newtheorem{lemma}{Lemma}

\let\emptyset\varnothing
\hypersetup{colorlinks = true, linkcolor=blue,urlcolor = blue,citecolor=blue}

\begin{document}

\title{A Theory of Covenant Accounting Adjustment\thanks{We thank Qi Chen, Judson Caskey, Elia Ferracuti, Henry Friedman, John Heater, Mark Kim, Volker Laux, Beatrice Michaeli, Daniel Saavedra, Katherine Schipper, Rahul Vashishtha, Mohan Venkatachalam, Hao Xue and workshop participants at DAR \& DART Accounting Theory Seminar, Duke, Southwest University of Finance and Economics of China, 2021 Stanford Accounting Summer Camp, UCLA, University of Melbourne and University of Waterloo for helpful comments of a previous version.}}
\author{Pingyang Gao\thanks{
		HKU Business School, University of Hong Kong. Email: pgao@hku.hk}, Xu Jiang\thanks{Fuqua School of Business, Duke University.  Email: xu.jiang@duke.edu}, and Jinzhi Lu\thanks{Department of Accountancy, City University of Hong Kong. Email: jinzhilu@cityu.edu.hk}}
\maketitle

\begin{abstract}
We develop an incomplete-contracting model with accounting-based covenants
to study how covenant accounting adjustments are made and what properties
they exhibit. Standard accounting rules (e.g., GAAP) can generate
false-alarm errors or undue-optimism errors. The manager can exert costly
effort to privately identify these errors and propose adjustments. If errors
are not corrected, control rights may be inefficiently allocated, leading to
costly renegotiation. We show that (1) adjustments always correct
false-alarm errors, but correct undue-optimism errors only when their
magnitude is small; and (2) the manager may expend socially wasteful effort
to identify these errors. The model yields testable empirical predictions
and policy implications.
\end{abstract}

\setlength{\baselineskip}{24pt} \setlength{\parindent}{15pt} \noindent Key
words: covenant accounting; information acquisition; incomplete contracting%
\newline
\newpage

\section{Introduction}

Covenant accounting adjustments—tailored modifications to GAAP
definitions, measurement rules, and recognition criteria—vary
substantially both across firms and within individual debt contracts. For
example, some agreements define adjusted EBITDA to include restructuring
charges and transaction fees (often subject to a cap), while others
explicitly prohibit these add-backs but permit non-cash impairment charges (%
\cite{dyreng2017direct}). Likewise, stock-based compensation is added back
in some contracts but expressly disallowed in others. Strikingly, even
within the same industry, two otherwise similar borrowers frequently employ
materially different adjustment sets.

Despite extensive empirical documentation of this heterogeneity, there is
surprisingly little theoretical research on how covenant accounting
adjustments emerge. Well-calibrated adjustments can materially improve
contracting efficiency, but identifying and negotiating them is costly.
Ideally, borrowers and lenders would jointly invest effort to uncover
adjustments that optimally balance efficiency gains against these costs, so
that the observed set would reflect the best feasible trade-off. In
practice, however, the parties differ systematically in their incentives and
capacity to identify value-enhancing adjustments. This asymmetry raises
several central questions: What motivates each party to incur the costs of
searching for adjustments? How do they strategically search, propose and
negotiate these modifications? And, most importantly, what are the
equilibrium properties of covenant accounting adjustments that emerge from
this bargaining process?

We study a model to answer these questions. In the model, the borrower (a
manager) and the lender enter into a debt covenant that allocates control
rights based on an accounting report. Our focus is on the negotiated rule
that produces the accounting report. We assume that the default rule (e.g.,
GAAP) is either perfect in reflecting the state or inaccurate. The
inaccuracy could be in either direction: ``false alarm" error (bad signal in
a good state) or ``undue optimism" error (good signal in a bad state).

In the benchmark case, both parties jointly search for inaccuracies, and any
information discovered becomes common knowledge. Subsequently, the manager
and the lender negotiate the contract terms, including whether to adopt an
accounting adjustment to correct the inaccuracy. Because correcting the
inaccuracy improves the ex-ante allocation of control rights and avoids
costly ex-post renegotiation, the resulting search effort achieves the
first-best level.

A key friction in practice, however, is that parties search for inaccuracies
privately and may strategically withhold their findings. We focus
specifically on the manager's incentives, as managers
typically possess superior information regarding the underlying transactions
and reporting choices that drive measurement error. To model this
informational advantage, we assume that after expending resources, the
manager receives a private signal regarding the error with some probability.
If the manager identifies an error, he can propose an adjustment or withhold
it by keeping silent. We assume that it is costless for the lender to verify
whether proposed adjustment is justified. An analogy to this assumption is
that it is difficult to solve a math problem, but it is much easier to
verify the answer.

Our main results show that the equilibrium property of covenant accounting
adjustments is asymmetric. When the manager dominates the process, the
adjustments always correct the false alarm error, but only correct the undue
optimism error when it is not significant. In other words, the covenant
accounting rules are more likely to contain undue optimism error than false
alarm error relative to GAAP.

The intuition behind this asymmetry is not immediately obvious, as the
manager faces a trade-off in both scenarios. Revealing and correcting the
false alarm error allows the manager to retain control more often; however,
because the lender anticipates receiving control less frequently, he demands
a higher interest rate to break even. Conversely, revealing and correcting
the undue optimism error forces the manager to surrender control more often,
but in exchange, the lender accepts a lower interest rate. Since both cases
involve balancing the retention of control rights against the cost of
borrowing, it is not initially clear why the equilibrium outcomes diverge.
The explanation lies in the dual effects of revealing and correcting the
error: First, it removes the information asymmetry between the manager and
the lender and redistributes the total surplus between two parties
(``redistribution effect''). Second, it increases the total surplus by
avoiding costly ex-post renegotiation (``production effect''). While the
second effect always benefits the manager, whether the first effect benefits
the manager depends on the direction of the error (i.e., false alarm or
undue optimism).

In the false alarm case, the manager—privately aware of the error%
—anticipates that the lender will obtain control with high
probability. However, if the error remains uncorrected, the lender accepts
the interest rate based on the belief that the manager may also be
uninformed, and thus assigns a lower probability to the false alarm error.
Since the lender underestimates the likelihood of obtaining control, the
manager expects the lender to earn a surplus strictly above the break-even
level. Disclosing the error removes the information asymmetry, and thus
reduces the lender's share of the total surplus. Because disclosing
simultaneously increases the total surplus and decreases the lender's share of it, the manager always discloses the false alarm
error. In other words, the redistribution effect and the production effect
work in the same direction.

In the undue optimism case, the manager—privately aware of the
error—anticipates that the lender will obtain control with low
probability. However, if the error remains uncorrected, the lender accepts
the interest rate based on the belief that the manager may also be
uninformed, and thus assigns a lower probability to the undue optimism
error. Since the lender overestimates the likelihood of obtaining control,
the manager expects the lender to earn a surplus strictly below the
break-even level. Disclosing the error removes the information asymmetry,
and thus increases the lender's share of the total surplus. Because
disclosing simultaneously increases the total surplus and increases the
lender's share of it, the manager faces a trade-off when
deciding whether to disclose the undue optimism error. In other words, the
redistribution effect and the production effect work in opposite directions.

Note that both effects increase with the magnitude of the undue optimism
error: the higher the magnitude, the more likely control rights are
misassigned ex-ante, resulting in more costly ex-post renegotiation. In
addition, with a higher magnitude, the manager understands that the lender
overestimates the likelihood of obtaining control to a larger degree. A main
insight of our result is that the wealth redistribution effect dominates the
production effect if and only if the magnitude of undue optimism is
sufficiently large.

Building on the main result, we find that the manager's incentive to
withhold the undue optimism error induces socially excessive effort in
identifying GAAP errors. The key mechanism is the option value of
concealment: because the manager can strategically suppress information that
is unfavorable to him, the private return to identifying the errors exceeds
the social return. While prior work documents a similar force in voluntary
disclosure settings (e.g., \cite{shavell1994acquisition}), we show that it
also arises in our setting with covenant accounting adjustments.

Our results have empirical and policy implications. Our main result explains
why accounting rules that generate false alarm errors are preferred to those
that generate undue optimism errors, as false alarm errors are more likely
to be found out and disclosed by the manager upfront, whereas the undue
optimism errors are more likely to be withheld. The results also imply that
we are more likely to see corrections of false alarm errors when adjusting
GAAP numbers for debt contracting, consistent with the usual practice to
exclude certain expense items (i.e., EBIT, EBITDA, or adjusted EBITDA). The
results also explain why it is sometimes better to commit to using the GAAP
rules even if such rules are prone to measurement errors, as such commitment
prevents managers' excessive effort of finding out whether the GAAP rules
are suitable. In other words, our results provide a justification for why
debt contracts sometimes exclude accounting changes.

\subsection{Related literature}

Our paper contributes to several streams of literature. First, it is related
to the broad empirical literature documenting the differences between
accounting rules used in the actual debt contract and standard rules such as
U.S. GAAP (e.g., \cite{leftwich1983accounting}, \cite{li2010negotiated}, and 
\cite{dyreng2017direct}). We are, to the best of our knowledge, the first to
provide an analytical explanation of why such a difference exists and what
type of difference is preferred. In particular, our results are consistent
with the findings in \cite{li2010negotiated} regarding asymmetric properties
of the adjustment of GAAP earnings in debt contracts. Section \ref{Section:
implications} provides more discussions of the connections between our model
and this stream of empirical literature. We respond to the call in \cite%
{armstrong2010role} and in \cite{christensen2016accounting} for more
research to ``investigate the financial reporting attributes that
debtholders value by examining the modifications to GAAP that are made in
the calculation of compliance with covenants'' and to adopt the
incomplete contracting theory approach to be useful in ``advancing our
knowledge of how accounting information affects contract efficiency''.

Our paper relates to analytical work on the optimal properties of earnings
in debt contracts, especially models with strategic managerial reporting.
Prior studies show that when managers can manipulate earnings, covenants can
become suboptimal under renegotiation (\cite{laux2022debt}) or be excluded
when manipulation and distortion costs outweigh their benefits (\cite%
{guttman2018debt}). In contrast, we study how GAAP can differ from
contractual earnings through managers' correction of GAAP
errors. As a result, covenants and renegotiation remain value-improving in
our setting, and our focus is on the asymmetry of false-alarm versus
undue-optimism errors, rather than on reporting precision. Our paper
complements \cite{laux2024accounting}: they show conservatism (i.e., false
alarm error) encourages manager's ex-post information acquisition. In
contrast, our results suggest that lenders prefer financial reporting rules
that produce false alarm errors rather than undue optimism errors, as false
alarm errors are more likely to be disclosed and corrected, whereas undue
optimism errors may not be.

Our paper relates to the literature on endogenously incomplete contracts and
socially wasteful information acquisition. Prior work shows that it can be
optimal to omit contingencies because they invite manipulation or
inefficient signaling (\cite{allen1992measurement}), are costly to enforce (%
\cite{glaeser2001reason}), or induce socially costly cognitive effort (\cite%
{tirole2009cognition}). Relatedly, in bargaining settings, parties may
overinvest in privately valuable information in ways that are socially
wasteful and can even prevent efficient trade (\cite{glode2012financial}; 
\cite{dang2008bargaining}). Building on incomplete-contracting approaches (%
\cite{aghion1992incomplete}), we bring these insights to debt contracting by
making measurement rules endogenous and studying how GAAP properties shape
managerial information acquisition and disclosure.

The paper is organized as follows. Section \ref{Section: the model} presents
the baseline model. Section \ref{Section: the equilibrium} provides some
preliminary analysis. Section \ref{sec:equilibrium} solves the equilibrium
of disclosure sub-game and presents the comparative statics analysis.
Section \ref{sec:extension} solves for the manager's equilibrium effort.
Section \ref{Section: implications} discusses empirical and policy
implications and connects to the empirical literature. Section \ref{Section:
conclusion} concludes. All proofs are in the appendix.

\section{The model \label{Section: the model}}

We adapt the incomplete contracting framework of \cite{aghion1992incomplete}
to include ex-ante negotiation of accounting rules. A penniless
borrower-manager seeks funding for the set-up costs $K$ of his new project
at date $0.$ We assume the lending market is competitive so the manager
makes a take-it-or-leave-it offer to a lender who accepts any offer that
allows him to break even.

The state $\theta$ is good ($G$) or bad ($B$) with equal probability. At
date $2$, the project is either continued ($a=1$) or restructured ($a=0$).
If it is continued in state $\theta $, the project pays out cash flow $Y$
with probability $\gamma _{\theta }$ and $0$ otherwise. If it is
restructured in state $\theta ,$ the project pays out cash flow $Y$ with
probability $\gamma _{\theta }$ and cash flow $0<y<Y$ with probability $%
1-\gamma _{\theta }$. In addition, continuation yields a private benefit $X$
to the manager. We assume $1 > \gamma_G > \gamma_B \geq 0$. The expected
joint surplus is: 
\begin{equation}
w(\theta, a) = \gamma_{\theta} Y + aX + (1-a)(1-\gamma_{\theta}) y.
\label{eqn: w(theta,a)}
\end{equation}

Although the state $\theta$ is observable by both parties, it is not
contractible. Instead, contracts rely on a date 2 accounting signal $r\in\{g,b\}$. The debt contract specifies a face value $D$ and allocates control rights based on $r$: the lender obtains control if and only if $r=b$. At date $2$, parties may renegotiate the allocation of control. We denote
the manager's bargaining power by $\tau \in [0,1]$ and assume renegotiation
consumes $\kappa \in \left(0,1\right) $ fraction of the joint surplus.

Parties begin with a default rule $R$ that generates the accounting signal $r
$. The interpretation of $R$ is flexible: it can represent standardized
reporting rules like GAAP, or simply the existing covenant definitions the
parties used in the previous year. Following \cite{aghion1992incomplete}, $R$
can be imperfect because it may not be perfectly tailored to the current
underlying transaction. Specifically, we consider two types of errors.

In the case of \textit{false alarm error}, the measurement rule satisfies 
\begin{equation*}
\Pr (r=b|G)=w\text{ and }\Pr (r=g|B)=0,
\end{equation*}
where $w\in(0,1)$ measures the degree of false alarm.

In the case of \textit{undue optimism error}, the measurement rule satisfies 
\begin{equation*}
\Pr (r=b|G)=0\text{ and }\Pr (r=g|B)=z,
\end{equation*}
where $z\in(0,1)$ measures the degree of undue optimism.

Because measurement errors can lead to inefficient ex-post allocation of
control rights, parties have an incentive to privately identify these flaws
and propose adjustments to the default rule $R$. This private information
search also creates strategic incentives: once a party identifies an error,
they may choose to either propose a correction or conceal the error,
depending on how the error affects their own expected payoff. Our main
departure from \cite{aghion1992incomplete} is the modeling of this
negotiation process— specifically incorporating
these strategic decisions regarding the correction of errors—and how parties eventually settle for an adjusted rule, denoted by $R^{\prime }$.	

In our baseline model, we focus on the manager's incentives to identify
measurement errors. To simplify the exposition, we introduce a new variable $%
x$ to characterize the degree of the measurement error in a continuous way.
We assume that $x$ has support $[-1,1]$ with density function $f(x)$. When $%
x\in \lbrack 0,1]$, the measurement rule has false alarm error with degree $x
$. When $x\in \lbrack -1,0]$, the measurement rule has undue optimism error
with degree $-x$. We assume that $f(x)$ is symmetric around 0 (i.e., $%
f(x)=f(-x)$ for any $x$) so that the distributions of the two types of
errors are a priori symmetric.

By exerting effort $p$, the manager can identify $x$ (i.e., becomes informed
about $x$) with probability $p$, at a cost of $c(p)$, where $c(p)$ increases
with $p$. We assume $c(p)$ is sufficiently convex to ensure that the effort
choice is interior (i.e., $p \in (0,1)$). If the manager identifies an
error, he can disclose it and propose an adjustment to the default
measurement rule. We use $R^{\prime }$ to denote the adjusted measurement
rule. As in \cite{tirole2009cognition}, once an adjustment is proposed, the
lender can verify its legitimacy at no cost. For simplicity, we assume $%
R^{\prime }$ generates a perfect measurement of the states.\footnote{%
Assuming $R^{\prime }$ is perfect significantly simplifies the analysis. Our
main insights hold as long as $R^{\prime }$ is more informative than $R$.}
If either the manager is uninformed or the manager is informed of $x$ but
chooses not to disclose, then the default $R$ will be used in contracting,
resulting in measurement errors. We use $d(x)\in \{0,1\}$ to denote the
informed manager's disclosure decision as a function of $x$, where $d=1$ ($%
d=0$) represents disclosing (withholding).

The timeline of the model is as follows:

\begin{itemize}
\item At date 0, the manager exerts effort $p$. With probability $p$, the
manager learns the error $x$ in the default measurement rule $R$.

\item At date 1, the manager decides whether to disclose the error and
propose the adjusted rule $R^{\prime }$. The parties adopt $R$ or $R^{\prime
}$ based on the manager's disclosure decision $d(x)$. The debt contract is
signed with face value $D$.

\item At date 2, the accounting signal $r$ is realized based on the chosen
measurement rule ($R$ or $R^{\prime }$). Control right is assigned to the
manager upon $r=g$ and to the lender upon $r=b$. State $\theta$ is realized
and observed by both parties. Possible renegotiation occurs and action $a$
(continuation or restructuring) is taken.

\item At date 3, all uncertainties are realized.
\end{itemize}

\subsection{Discussion of model assumptions}

We assume the manager cannot commit to not acquiring information about the
errors in $R$. This reflects the reality that a manager's information
acquisition is inherently unobservable and therefore non-contractible.
Furthermore, because the manager is intimately involved in the firm's daily
operations, they will inevitably encounter evidence regarding the
suitability of the default rule, making any promise to refrain from learning
impossible to enforce.

We assume the lender has no ex-ante bargaining power and thus no incentive
to acquire private information about the errors. This model feature is not
crucial for our insights. If the lender has all the bargaining power, then
the main result will be symmetric: the adjustments always correct the undue
optimism error, but only correct the false alarm error when it is not
significant. When both parties have bargaining power and can acquire
information about the errors, our main insights remain intact: the manager
will only propose adjustments to false alarm error, and the lender will only
propose adjustments to undue optimism error, whereas the aggressiveness of
parties' disclosures depend on the magnitude of the errors and the
bargaining power.

\section{Preliminary analysis \label{Section: the equilibrium}}

\subsection{The optimal allocation of control rights}

We make two technical assumptions regarding the exogenous parameters to
ensure that the control rights allocation rule does not result in trivial
solutions.

\begin{assumption}
\label{assumption1} $L_B\equiv(1-\gamma _{B})y-X>0$, $L_G\equiv X-\left( 
1-\gamma _{G}\right) y>0$, and $L_G>L_B$.
\end{assumption}

\begin{assumption}
\label{assumption2} $K>y$.
\end{assumption}

Assumption \ref{assumption1} implies that the socially optimal action
depends on the state. Specifically, restructuring is optimal in the bad
state ($\theta=B$) to capture the liquidation value $y$, while continuation
is optimal in the good state ($\theta=G$) to preserve the private benefit $X$%
. We define the efficiency losses from deviating from the first-best action
in each state as: 
\begin{align}
L_{B} &\equiv w(B,0)-w(B,1)=\left( 1-\gamma _{B}\right) y-X>0,  \label{L_B}
\\
L_{G} &\equiv w(G,1)-w(G,0)=X-\left( 1-\gamma _{G}\right) y>0.  \label{L_G}
\end{align}
Consequently, the first-best action is $a_{G}^{FB}=1$ and $a_{B}^{FB}=0$.

The project's total expected payoff is divided between the manager and the
lender under the debt contract $D.$ For a given face value, the lender's
share of the project's expected payoff in state $\theta $ with action $a$,
denoted as $w^{L}(\theta ,a)$, is 
\begin{equation*}
w^{L}(\theta ,a)=\gamma _{\theta }D+\left( 1-a\right) \left( 1-\gamma
_{\theta }\right) \min \{D,y\}.
\end{equation*}

Even though the value $D$ will be endogenously determined in equilibrium, we
can infer beforehand that $D\geq K.$ Otherwise, the lender cannot recoup the
principal $K$. Assumption \ref{assumption2} then implies that $D>y.$ As a
result, $w^{L}(\theta ,a)$ is simplified as%
\begin{equation*}
w^{L}(\theta ,a)=\gamma _{\theta }D+\left( 1-a\right) \left( 1-\gamma
_{\theta }\right) y.
\end{equation*}%
It can be verified that $w^{L}(\theta ,0)-w^{L}(\theta ,1)=\left( 1-\gamma
_{\theta }\right) y>0.$ Thus, under Assumption \ref{assumption2}, the lender
always prefers restructuring, regardless of the state.

Similarly, the manager's share of the project's expected payoff in state $%
\theta $ with action $a$, denoted as $w^{M}(\theta ,a)$, is 
\begin{equation*}
w^{M}(\theta ,a)=w\left( \theta ,a\right) -w^{L}\left( \theta ,a\right)
=\gamma _{\theta }(Y-D)+aX.
\end{equation*}
It is straightforward to show that $w^{M}(\theta ,1)-w^{M}(\theta ,0)=X>0.$
Thus, under Assumption \ref{assumption2}, the manager always prefers
continuation, regardless of the state.

Collecting these results, we have the following lemma, with its proof
already laid out above.

\begin{lemma}
\label{Lemma: conflict of interest}Under Assumptions \ref{assumption1} and %
\ref{assumption2}, the socially optimal action is to continue the project if
and only if the state is good. However, regardless of the state, the manager
prefers continuation while the lender prefers restructuring.
\end{lemma}

When there is no accounting information, the total expected payoff under
continuation versus restructuring is $L=\frac{1}{2} L_G -\frac{1}{2}L_B$.
Thus, under Assumption \ref{assumption1} (i.e., $L_G>L_B$), the optimal
action is continuation in the absence of information.\footnote{
This assumption is not crucial for our results. If $L_G<L_B$, the optimal 
action is restructuring in the absence of information. Our results hold as 
long as the accounting signal can change the optimal action.}

When the default measurement rule $R$ incurs error, the accounting signal is
a noisy measure of the state. Let $\mu $ denote the posterior probability of
the good state. The total expected payoff under continuation versus
restructuring is expressed as $L\equiv \mu L_{G}-(1-\mu )L_{B}$. Conditional
on $r=g$, the good state is more likely, implying $\mu >\frac{1}{2}$. Under
Assumption \ref{assumption1} (i.e., $L_{G}>L_{B}$), we have $L>0$, and it is
socially optimal to allocate the control rights to the manager. Conditional
on $r=b$, the bad state is more likely, implying $\mu <\frac{1}{2}$. If $\mu 
$ is not very low conditional on $r=b$, then $L>0$ conditional on $r=b$. In
this case, even $r=b$ does not change the optimal action. Thus, it is always
optimal to allocate the control rights to the manager, resulting in a
trivial case.

Therefore, from now on, our analysis focuses on the case where $\mu$ is
sufficiently low conditional on $r = b$. Consequently, it is socially
optimal to allocate the control right to the manager when $r = g$ and to the
lender when $r = b$.

\subsection{First-best benchmark}

In the first-best benchmark, any information that the manager obtains is
observed by both parties to correct for possible measurement errors and thus
avoid any possible social loss from renegotiation. Recall that renegotiation
happens with a probability of $\frac{1}{2}x$ in the case of false alarm
error and with a probability of $-\frac{1}{2}x$ in the case of undue
optimism error. Hence, the measurement errors, if uncorrected, reduce the
total surplus by 
\begin{align*}
L=& \frac{1}{2}\int_{0}^{1}\kappa L_{G}xf(x)dx-\frac{1}{2}
\int_{-1}^{0}\kappa L_{B}(-x)f(x)dx \\
=& \frac{1}{2}\int_{0}^{1}(\kappa L_{G}+\kappa L_{B})xf(x)dx.
\end{align*}
Exerting effort $p$ will increase the total surplus by $pL$, as the errors
are corrected with probability $p$. Thus, the first-best level of effort
(denoted by $p^{FB}$) satisfies 
\begin{equation}
c'(p^{FB})=\frac{1}{2}\int_{0}^{1}(\kappa L_{G}+\kappa L_{B})xf(x)dx 
\text{.}  \label{FOC}
\end{equation}
Obviously, $p^{FB}$ increases in $\kappa$, $L_G$, and $L_B$, due to the
convexity of the cost function $c(.)$.

\section{Main results \label{sec:equilibrium}}

We solve the model through backward induction. In this section, we solve for
the manager's disclosure strategy given the effort choice. We then solve for
the manager's equilibrium effort choice in Section \ref{sec:extension}.

\subsection{The manager's error disclosure strategy}

Denote the manager's and the lender's payoff when the state is $\theta $ and
the accounting signal is $r$ as $u_{\theta}^{r}$ and $v_{\theta }^{r}$,
respectively. Denote $w_{\theta }^{r}$ as the social payoff, which is the
sum of $u_{\theta}^{r}$ and $v_{\theta }^{r}$.

Table \ref{Table:fg_t=2} below lists the payoff for the manager and the
lender at $t=2$ when there is false alarm error. 
\begin{table}[H]
\centering 
\begin{tabular}{|l|l|l|l|l|}
\hline
$\theta $ & $r$ & $u_{\theta }^{r}$ & $v_{\theta }^{r}$ & $w_{\theta }^{r}$
\\ \hline
$G$ & $g$ & $\gamma _{G}(Y-D)+X$ & $\gamma _{G}D$ & $\gamma _{G}Y+X$ \\ 
$G$ & $b$ & $\gamma _{G}(Y-D)+\tau (1-\kappa )L_{G}$ & $\gamma
_{G}D+(1-\gamma _{G})y+\left( 1-\tau \right) (1-\kappa )L_{G}$ & $\gamma
_{G}Y+X-\kappa L_{G}$ \\ 
$B$ & $b$ & $\gamma_B(Y-D)$ & $\gamma_BD+(1-\gamma_B)y$ & $%
\gamma_BY+(1-\gamma_B)y$ \\ \hline
\end{tabular}%
\caption{Payoff for the manager and the lender at t=2 with false alarm errors
}
\label{Table:fg_t=2}
\end{table}
When $\theta =G$ and $r=g$, the manager has the control rights and
continuation is the socially efficient decision, resulting in no
renegotiation. Since $D \le Y$, the manager gets the expected cash flow, $%
\gamma _{G}(Y-D)$, plus the private benefit $X$, whereas the lender gets the
expected cash flow, $\gamma _{G}D$. Similarly, when $\theta =B$ and $r=b$,
the lender has the control rights and restructuring is the socially
efficient decision, again resulting in no renegotiation, the lender getting $%
\gamma_BD+(1-\gamma_B)y$ and the manager getting $\gamma_B(Y-D)$ as $y<D$.
When $\theta =G$ and $r=b$, the lender has the control rights but
continuation is the socially optimal decision. Therefore there is
renegotiation and the project will be continued. The manager gets the
expected cash flow from restructuring, $\gamma _{G}(Y-D)$, plus his share of
the surplus from renegotiation, i.e., $\tau (1-\kappa )L_{G}$. Similarly,
the lender gets the expected cash flow from restructuring, $\gamma
_{G}D+(1-\gamma _{G})y$, plus her share of the surplus from renegotiation, $%
\left( 1-\tau \right) (1-\kappa )L_{G}$.

Table \ref{Table:fb_t=2} below lists the payoff for the manager and the
lender at $t=2$ when there is undue optimism error.

\begin{table}[H]
\centering
\begin{tabular}{|l|l|l|l|l|}
\hline
$\theta $ & $r$ & $u_{\theta }^{r}$ & $v_{\theta }^{r}$ & $w_{\theta }^{r}$
\\ \hline
$G$ & $g$ & $\gamma _{G}(Y-D)+X$ & $\gamma _{G}D$ & $\gamma _{G}Y+X$ \\ 
$B$ & $g$ & $\gamma_B(Y-D)+X+\tau (1-\kappa )L_{B}$ & $\gamma_BD+\left(
1-\tau \right) (1-\kappa )L_{B}$ & $\gamma_BY+X+(1-\kappa)L_B$ \\ 
$B$ & $b$ & $\gamma_B(Y-D)$ & $\gamma_BD+(1-\gamma_B)y$ & $%
\gamma_BY+(1-\gamma_B)y$ \\ \hline
\end{tabular}%
\caption{Payoff of the manager and the lender at t=2 with undue optimism
errors}
\label{Table:fb_t=2}
\end{table}
The first row and the third row are the same as the false alarm error case.
When $\theta =B$ and $r=g$, the manager has the control rights but
restructuring is the socially optimal decision. Therefore there is
renegotiation and the project will be restructured. The manager gets the
expected payoff from continuation, $\gamma _{B}(Y-D)+X$, plus the manager's
share of the surplus from renegotiation, i.e., $\tau (1-\kappa )L_{B}$.
Similarly, the lender gets the expected cash flow from continuing, $\gamma
_{B}D$, plus the lender's share of the surplus from renegotiation, $\left(
1-\tau \right) (1-\kappa )L_{B}$.

The next proposition presents our first main result.

\begin{proposition}
\label{prop:D} In equilibrium, the manager always discloses false alarm
errors.
\end{proposition}

The intuition of Proposition \ref{prop:D} lies in the dual effects of
revealing and correcting the error. First, it removes the information
asymmetry between the manager and the lender and redistributes the total
surplus between the two parties (\textquotedblleft redistribution
effect\textquotedblright ). Second, it increases the total surplus by
avoiding costly ex-post renegotiation (\textquotedblleft production
effect\textquotedblright ). While the second effect always benefits the
manager, whether the first effect benefits the manager depends on the
direction of the error.

In the false alarm case, the manager—privately aware of the error%
—anticipates that the lender will obtain control with high
probability. However, if the error remains uncorrected, the lender accepts
the interest rate based on the belief that the manager may also be
uninformed, and thus assigns a lower probability to the false alarm error.
Since the lender underestimates the likelihood of obtaining control, the
manager expects the lender to earn a surplus strictly above the break-even
level. Disclosing the error removes the information asymmetry, and thus
reduces the lender's share of the total surplus. Because disclosing
simultaneously increases the total surplus and decreases the lender's share of it, the manager always discloses the false alarm
error. In other words, the redistribution effect and the production effect
work in the same direction.

In the undue optimism case, the manager—privately aware of the
error—anticipates that the lender will obtain control with low
probability. However, if the error remains uncorrected, the lender accepts
the interest rate based on the belief that the manager may also be
uninformed, and thus assigns a lower probability to the undue optimism
error. Since the lender overestimates the likelihood of obtaining control,
the manager expects the lender to earn a surplus strictly below the
break-even level. Disclosing the error removes the information asymmetry,
and thus increases the lender's share of the total surplus. Because
disclosing simultaneously increases the total surplus and increases the
lender's share of it, the manager faces a trade-off when
deciding whether to disclose the undue optimism error. In other words, the
redistribution effect and the production effect work in opposite directions.

The next proposition presents the result regarding the manager's correction
of the undue optimism error.

\begin{proposition}
\label{prop:threshold} In equilibrium, the manager discloses $x$ if and only
if $x>x^{\ast }$, where $x^{\ast }\in\lbrack-1,0)$. In other words, the
manager always discloses false alarm errors, and discloses undue optimism
errors if and only if their magnitude is sufficiently small. The threshold $%
x^*$ is determined by the following equation: 
\begin{equation}
pC_{1}\int_{-1}^{x^{\ast }}F(x)dx+p(C_{2}-C_{1})x^{\ast }F(x^{\ast
})+(1-p)(C_{2}x^{\ast }-C)=0\text{,}  \label{eqn:indiffmain}
\end{equation}
where $C$, $C_{1}$, and $C_{2}$ are defined in \eqref{eqn:C}, \eqref{eqn:C1}%
, and \eqref{eqn:C2} in the appendix.
\end{proposition}

As discussed above, the manager faces a trade-off between production effect
and redistribution effect when deciding whether to disclose the undue
optimism error. According to Table \ref{Table:fb_t=2}, when the manager
reveals the undue optimism error $x$, the production effect increases
expected social surplus by $|x|\kappa L_{B}$, while the redistribution
effect increases the lender's share by $K-(|x|v_{B}^{g}+(1-|x|)v_{B}^{b})$,
which is equivalent to $(v_{B}^{b}-v_{B}^{g})|x|+K-v_{B}^{b}$.

Notably, both effects are linear and increasing in the magnitude of the
undue optimism error. The manager compares these effects to decide whether
to disclose $x$. If the renegotiation cost ($\kappa$) is sufficiently large
that the production effect always dominates, we obtain a corner solution
where the manager always discloses $x$. Otherwise, if the renegotiation cost
is moderate, the slope of the redistribution effect is steeper.
Consequently, the redistribution effect dominates if and only if the
magnitude of $x$ is sufficiently large, a result driven by the intercept
term in $(v_B^g-v_B^b)|x|+K-v_B^b$. Intuitively, this intercept ($K-v_B^b$)
is negative because, when the state is bad and $r=b$, the lender obtains
control and anticipates a payoff exceeding the break-even level, $K$.

Equation \eqref{eqn:indiffmain} is obtained by imposing the condition that
the marginal type manager (i.e., $x^{\ast }$) must be indifferent between
disclosing and withholding. In the appendix, we also show that the solution
to \eqref{eqn:indiffmain} is unique when $\kappa $ is small. Intuitively,
for the equilibrium to be unique, the manager's incentive to withhold must
increase with $x^*$. This condition holds precisely when $\kappa$ is small,
as the redistribution effect dominates the production effect.

\subsection{Comparative statics of the disclosure threshold\label{sec:cs}}

We now conduct comparative statics analysis to examine how $x^{\ast }$
varies with the exogenous parameters. First, we analyze how the parameters $%
C_{1}$, $C_{2}$ and $C$ in equation \eqref{eqn:indiffmain} change with
respect to exogenous parameters. Note that the parameter 
\begin{equation*}
C_{1}\equiv \frac{1}{2}(v_{B}^{b}-v_{B}^{g})=\frac{1}{2}((1-\gamma
_{B})y-(1-\tau )(1-\kappa )L_{B})>0,
\end{equation*}
captures the lender's marginal loss when the degree of undue optimism
increases. The parameter 
\begin{equation*}
C_{2}=\frac{1}{2}(u_{B}^{g}-u_{B}^{b})=\frac{1}{2}(X+\tau (1-\kappa
)L_{B})>0,
\end{equation*}
captures the manager's marginal gain when the degree of undue optimism
increases.

To see the meaning of $C$, we further define a third parameter 
\begin{equation*}
C_{3}=v_{G}^{b}-v_{G}^{g}=\frac{1}{2}((1-\gamma _{G})y+(1-\tau )(1-\kappa
)L_{G})>0,
\end{equation*}%
which captures the lender's marginal gain when the degree of false alarm
increases. The parameter $C$ in equation \eqref{eqn:indiffmain} can be
expressed as 
\begin{equation*}
C=(C_{3}-C_{1})\int_{0}^{1}xf(x)dx.
\end{equation*}

For the rest of the analysis, we focus on the case where $\kappa$ is
sufficiently small so that the disclosure threshold is interior. We can thus
ignore the effect of the second term in \eqref{eqn:indiffmain} (i.e., $%
p(C_{2}-C_{1})x^{\ast }F(x^{\ast })$) when deriving the comparative statics,
as $C_{2}-C_{1}=-\kappa L_{B}$.

Differentiating \eqref{eqn:indiffmain} with respect to any parameter $\beta$
leads to 
\begin{align*}
\frac{\partial pC_1}{\partial \beta}\int_{-1}^{x^*}F(x)dx+pC_1F(x^*)\frac{%
\partial x^*}{\partial \beta}+ (1-p)C_2\frac{\partial x^*}{\partial \beta}%
+x^*\frac{\partial(1-p)C_2}{\partial \beta}-\frac{\partial (1-p)C}{\partial
\beta}=0.
\end{align*}
Solving for $\frac{\partial x^*}{\partial \beta}$ leads to 
\begin{align}
\frac{\partial x^*}{\partial \beta}&= \frac{\frac{\partial (1-p)C}{\partial
\beta}-\frac{\partial pC_1}{\partial \beta}\int_{-1}^{x^*}F(x)dx-x^*\frac{%
\partial(1-p)C_2}{\partial \beta}}{pC_1F(x^*)+(1-p)C_2}  \notag \\
&\propto \frac{\partial (1-p)C}{\partial \beta}-\frac{\partial pC_1}{%
\partial\beta}\int_{-1}^{x^*}F(x)dx-x^*\frac{\partial(1-p)C_2}{\partial \beta%
}.  \label{CS}
\end{align}
Using the equation $C=(C_3-C_1)\int_0^1 xf(x)dx$, we can rewrite \eqref{CS}
as 
\begin{align}
\frac{\partial x^*}{\partial \beta}&\propto \frac{\partial (1-p)C_3}{%
\partial \beta}\int_0^1 xf(x)dx-(\frac{\partial pC_1}{\partial\beta}%
\int_{-1}^{x^*}F(x)dx+\frac{\partial (1-p)C_1}{\partial \beta}\int_0^1
xf(x)dx)-x^*\frac{\partial(1-p)C_2}{\partial \beta}.  \label{CSalterna}
\end{align}

To understand \eqref{CSalterna}, note that upon non-disclosure, the lender
rationally believes that either the manager failed to discover $x$ or chose
to withhold it. We have established in Proposition \ref{prop:threshold} that
the manager always discloses false alarm errors. Thus, false alarm errors
reduce the lender's payoff in the first case, whereas undue optimism errors
increase the lender's payoff in both cases. When $C_{3}$ increases or when $%
C_{1}$ decreases, the lender's expected payoff conditional on non-disclosure
becomes higher and thus demands a lower interest rate to break-even. This
increases the manager's incentive to withhold undue optimism errors. As a
result, the first term of \eqref{CSalterna} is positive when $\beta =C_{3}$,
and the second term of \eqref{CSalterna} is negative when $\beta =C_{1}$.
Finally, an increase in $C_{2}$ increases the manager's incentive to
withhold undue optimism errors. Thus, the third term of \eqref{CSalterna} is
positive when $\beta =C_{2}$.

Recall that our model has the following exogenous parameters: $\gamma _{G}$, 
$\gamma _{B}$, $y$, $\tau $, $\kappa $, $X$, $p$. The next table summarizes
the effects of those parameters on $C_{1}$, $C_{2}$, $C_{3}$. 
\begin{table}[H]
\centering
\begin{tabular}{|l|l|l|l|}
\hline
\text{} & $C_1$ & $C_2$ & $C_3$ \\ \hline
$\gamma_G$ & 0 & 0 & - \\ \hline
$\gamma_B$ & - & - & 0 \\ \hline
$y$ & + & + & + \\ \hline
$X$ & 0 & + & + \\ \hline
$\tau$ & + & + & - \\ \hline
$\kappa$ & + & - & - \\ \hline
\end{tabular}%
\caption{The effect of parameters on $C$, $C_{1}$, $C_{2}$, and $C_{3}$}
\label{Table: CS1}
\end{table}

In order to understand the underlying intuition for the comparative statics,
we need to understand how $C_{1}$, $C_{2}$ and $C$ vary with the exogenous
parameters.

We first look at how $C_{1}$ varies with the exogenous parameters. Since $%
C_{1} $ captures the lender's marginal loss when the degree of undue
optimism error increases, it is not affected by $\gamma _{G}$. The lender
suffers less loss when the project has a higher probability of success or
when the restructuring value becomes lower. This explains $\frac{\partial
C_{1}}{\partial \gamma _{B}}<0$ and $\frac{\partial C_{1}}{\partial y}>0$.
In addition, a higher $X$ implies that the value gain from restructuring
becomes lower, as $L_{B}$ becomes lower. As a result, the lender suffers
more from undue optimism when $X$ becomes higher (i.e., $\frac{\partial C_{1}%
}{\partial X}>0$). Finally, it is straightforward that $C_{1}$ increases
with $\tau $ and $\kappa $, as increases in those parameters reduce the
lender's expected payoff (and thus increase the lender's marginal loss) as
either the lender's share or the total surplus decreases.

We next look at how $C_{2}$ varies with the exogenous parameters. Since $%
C_{2}$ captures the manager's marginal benefit when the degree of undue
optimism error increases, it is not affected by $\gamma _{G}$. A higher $y$
or a lower $\gamma _{B}$ implies that the manager's gain from renegotiation
becomes higher, as $L_{B}$ becomes higher. This explains $\frac{\partial
C_{2}}{\partial y}>0$ and $\frac{\partial C_{2}}{\partial \gamma _{B}}<0$.
Finally, it is straightforward that $C_{2}$ increases with $\tau$ and $X$,
as an increase in these parameters raises the manager's expected payoff (and
thereby the manager's marginal benefit). Conversely, $C_{2}$ decreases with $%
\kappa$, as an increase in $\kappa$ reduces the total surplus, thereby
lowering the manager's expected payoff.

Finally, we look at how $C_{3}$ varies with the exogenous parameters. Since $%
C_{3}$ captures the lender's marginal benefit when the degree of false alarm
error increases, it is not affected by $\gamma _{B}$. The lender enjoys more
benefit when the project has a lower probability of success, as the lender
is more likely to be in control, or when the restructuring value becomes
higher, as the restructuring value accrues more to the lender. In addition,
a higher $X$ implies that the value gain from restructuring becomes higher,
as $L_{G}$ becomes higher. Thus, the lender enjoys more benefit from false
alarm error when $X$ becomes higher (i.e., $\frac{\partial C_{3}}{\partial X}%
>0$). Finally, it is straightforward that $C_{3}$ decreases with $\tau $ and 
$\kappa $, for the same reason as that of $C_{1}$.

To sum up, each parameter can affect $x^{\ast }$ through the three channels:
(i) the lender's loss due to undue optimism error, (ii) the manager's
marginal gain due to undue optimism error, and (iii) the lender's gain due
to undue optimism error. The next proposition presents the comparative
statics result.

\begin{proposition}
\label{prop:cs} The comparative statics of $x^{\ast }$ with respect to
general symmetric distributions and the special case of the uniform
distribution are summarized in the following two tables: 
\begin{table}[H]
\centering
\begin{tabular}{|l|l|l|l|l|l|l|l|}
\hline
& $\gamma_G$ & $\gamma_B$ & $y$ & $X$ & $\tau$ & $\kappa$ & $p$ \\ \hline
$x^*$ & - & ? & ? & ? & ? & - & - \\ \hline
\end{tabular}%
\caption{The effect of parameters on $x^{\ast }$ under general symmetric
distribution of $x$, ``?'' means ambiguous effect}
\label{Table:CS2}
\end{table}

\begin{table}[H]
\centering
\begin{tabular}{|l|l|l|l|l|l|l|l|}
\hline
& $\gamma_G$ & $\gamma_B$ & $y$ & $X$ & $\tau$ & $\kappa$ & $p$ \\ \hline
$x^*$ & - & + & - & + & - & - & - \\ \hline
\end{tabular}%
\caption{The effect of parameters on $x^*$ assuming uniform distribution of $%
x$}
\label{Table:CS3}
\end{table}
\end{proposition}

Table \ref{Table:CS2} presents the comparative statics under the general
distribution. Some predictions are ambiguous (denoted by \textquotedblleft
?\textquotedblright ) because the three channels may work in different
directions. Table \ref{Table:CS3} presents the comparative statics under the
uniform distribution of $x$. For all parameters except $X$, the first
channel (i.e., the lender's loss due to undue optimism) dominates the other
two, as this is the main case when the lender would like to price protect,
which has the strongest effect on the manager's payoff and thus disclosure
incentive. For the manager's private benefit ($X$), the second channel
(i.e., the manager's marginal gain due to undue optimism) dominates the
other two, as the private benefit results in the manager being more likely
to continue, which is further exacerbated by the undue optimism error. Thus,
the manager's marginal gain due to the undue optimism error has a dominant
effect on the manager's incentive to disclose. Finally, similar to the
intuition from the prior literature (e.g., \cite{dye85jar} and \cite%
{jung1988disclosure}), the manager becomes more likely to disclose $x$ when
he is more likely to be informed (i.e., $\frac{\partial x^{\ast }}{\partial p%
}<0$).

\section{The manager's effort choice\label{sec:extension}}

Having established the equilibrium of the disclosure sub-game, we solve for
the manager's equilibrium effort and compare it with the first-best level.
By exerting effort $p$, the manager's expected payoff is a weighted average
of his payoff in three cases: (i) the manager is uninformed (an event
denoted by $\emptyset$); (ii) the manager is informed and $x>x^*$, in which
case he discloses $x$; and (iii) the manager is informed and $x<x^*$, in
which case he withholds $x$. Thus, the manager's payoff from exerting effort 
$p$ can be written as: 
\begin{align*}
E[u]=(1-p)E[u|\emptyset]+p\int_{-1}^{x^*}E[u|x, d=0]f(x)dx+p
\int_{x^*}^{1}E[u|x, d=1] f(x)dx.
\end{align*}
The first order condition that determines the optimal effort $p^*$ is 
\begin{align}
c'(p^*)&=\int_{-1}^{x^*}E[u|x, d=0]f(x)dx+
\int_{x^*}^{1}E[u|x,d=1]f(x)dx-E[u|\emptyset]  \label{pstar}
\end{align}

Details for the right-hand side of \eqref{pstar} are provided in the
appendix. The next proposition compares $p^*$ with $p^{FB}$.

\begin{proposition}
\label{prop:effort} The manager over-invests in information acquisition: $
p^{\ast }>p^{FB}$.
\end{proposition}

The intuition of this proposition is similar to that in the prior literature
(e.g., \cite{shavell1994acquisition}) that when the (potentially) informed
party can strategically disclose after exerting effort to acquire
information, the informed party tends to exert socially excessive effort due
to the option to withhold information when such information would harm the
informed party. In our setting, the information is related to the
suitability of the current measurement rule. The option to withhold such
information when the measurement rule generates undue optimism error results
in the manager exerting excessive effort.

\section{Relation to the empirical literature and implications\label%
{Section: implications}}

As discussed in the introduction, our results is related to the empirical
literature on the adjustment of GAAP earnings numbers in actual debt
contracts. \cite{leftwich1983accounting} is probably among the first to
empirically document such a difference. \cite{leftwich1983accounting}
provides evidence consistent with the explanation that adjustments from U.S.
GAAP is meant to prevent managers from engaging in behaviors that exploit
debtholders. This explanation is also consistent with our finding that GAAP
precludes undue optimism errors to prevent managers from withholding them. 
\cite{leftwich1983accounting} also argues that debtholders rely on
\textquotedblleft price protection\textquotedblright\ and so may be
indifferent to the terms of the debt issue but such \textquotedblleft price
protection\textquotedblright\ will affect shareholders' preferences. In our
model, it is indeed the case. Debtholders rely on such \textquotedblleft
price protection\textquotedblright\ to break even whereas such
\textquotedblleft price protection\textquotedblright\ affects the manager's
disclosure decisions and thus the manager's preference to different
accounting rules.

More recently, \cite{li2010negotiated} studies the adjustments of net income
and net worth in private debt contracts. He finds that transitory earnings
are usually excluded. To the extent that transitory earnings merely add
noise and is equally likely to generate false alarm errors and undue
optimism errors, it is not directly related to our predictions.
Nevertheless, Li's finding that there is not much conservative adjustment in
the debt contract is generally consistent with our results as we predict
that the GAAP rules should generate more false alarm errors relative to
undue optimism errors so there is not much additional conservative
adjustment left.

Our results provide the following implications. First, there are more
corrections for false alarm errors than corrections of undue optimism
errors, implying subsequently there will be more renegotiations due to undue
optimism errors. Our results provide a justification for accounting rules
that generate more false alarm errors relative to undue optimism errors when
errors are unavoidable: the managers have strong incentives to find out and
subsequently disclose false alarm errors relative to undue optimism errors.

Second, our comparative statics results show that withholding of undue
optimism errors would be more severe if the firm is less profitable,
renegotiation is less costly and the conflict of interest between
debtholders and shareholders is sufficiently large. We thus predict that the
GAAP rules would be more likely to exhibit false alarm errors when the
number of lenders in a syndicated loan is smaller or for private versus
public debt (as renegotiation cost will be smaller); and for debt contract
with more earnings-based covenants (as the conflict of interest will be
higher). To the extent that more false alarm errors imply more conservative
accounting, the latter prediction is consistent with \cite{nikolaev2010debt}.

Third, the correction of accounting errors in the debt contracts is more
likely to be correcting for false alarm errors (e.g., excluding certain
expense items), which is consistent with anecdotal evidence that EBIT,
EBITDA or adjusted EBITDA are often used to replace GAAP earnings in debt
contracts. To the extent that more false alarm errors imply more
conservative accounting, our results can be viewed as providing a
reconciliation between why GAAP earnings are conservative and adjusted
earnings numbers usually undo such conservatism, as documented in \cite%
{dyreng2017direct}. Conservative accounting, by reducing undue optimism
error, reduces the chances that managers withhold such error. Managers would
more often disclose false alarm errors and then adjust the accounting
numbers correspondingly to correct for such errors.

Finally, we provide an explanation on why debt contract may exclude
accounting method changes. Empirically, \cite{beatty2002importance} find
that in some debt contracts, lenders exclude both voluntary and mandatory
accounting changes in the calculation of covenant compliance and charge a
lower interest rate in return. They interpret this as the value of the
borrower's \textquotedblleft accounting flexibility\textquotedblright. In
our model, such \textquotedblleft accounting flexibility\textquotedblright\
stems from the borrower's excessive incentive to search for information
related to accounting changes. Since such a search can be socially wasteful,
the lender may find it optimal to exclude accounting changes to discourage
the search.

\section{Conclusion \label{Section: conclusion}}

We study how borrowers and lenders negotiate accounting measurement rules in
debt contracts to understand covenant design when GAAP is imperfect. We show
that managers' incentives to disclose GAAP inaccuracies
are asymmetric: they always disclose and correct false-alarm errors but tend
to withhold undue-optimism errors. This strategic concealment leads managers
to exert (socially) excessive effort to identify GAAP inaccuracies.

Our results provide an explanation of why the GAAP numbers are more likely
to be corrected for false alarm errors in debt contracting, and also why
sometimes contracts exclude all changes to the GAAP rules. We hope future
research can build on our framework to provide further insights into
covenant accounting adjustments. In particular, it would be valuable to
explore how these adjustments relate to managerial incentives, such as
earnings manipulation.

\newpage \appendix
\section{Proofs}

We first introduce some notations. We use $u$ to denote the manager's payoff
and $v$ to denote the lender's payoff. We use $t$ to denote the manager's
type and $m$ to denote the manager's message. Both $t$ and $m$ come from the
set $T$ of all possible types: $T=\{x|x\in[-1,1]\}\cup \emptyset$, where $%
\emptyset$ stands for the uninformed type. The manager's disclosure strategy 
$d(.)$ maps a type $t\in T$ to a message $m\in T$. The informed types can
imitate the uninformed types but not vice-versa. Therefore, the possible
strategies are: $d(\emptyset)=\emptyset$, $d(x)\in\{x,\emptyset\}$. For
simplicity, we use $d(x)\in \{0,1\}$ to denote the informed manager's
disclosure decision as a function of $x$, where $d=1$ ($d=0$) represents
disclosing (withholding).

\subsection{Proof of Proposition \ref{prop:D}}

\subsubsection{Step 1: Prove that the equilibrium must be a threshold
	strategy}

First, consider the case of $x<0$ (undue optimism). When the manager discloses $%
x $, his expected payoff is 
\begin{align*}
	E[u|x,d=1]=\frac{1}{2}(\gamma_G(Y-D_1)+X)+\frac{1}{2}(\gamma_B(Y-D_1)),
\end{align*}
where $D_1$ is the face value of the debt when $d=1$.

When the manager withholds $x$, his payoff is 
\begin{align*}
	E[u|x,d=0]=&\;\frac{1}{2}(\gamma_G(Y-D_0)+X)+\frac{1}{2}(1+x)\gamma_B(Y-D_0)
	\\
	&-\frac{1}{2}x(\gamma_B(Y-D_{0})+X+\tau (1-\kappa )L_{B}),
\end{align*}
where $D_0$ is the face value of the debt when $d=0$.

Therefore, the incremental payoff of disclosing relative to withholding,
denoted by $\Delta(x)$, is 
\begin{align}  \label{eqn:incremental1}
	\Delta(x)=\frac{1}{2}(\gamma_G+\gamma_B)(D_0-D_1)+\frac{1}{2}x(X+\tau
	(1-\kappa )L_{B}).
\end{align}

Second, consider the case of $x>0$ (false alarm). When the manager discloses $x$%
, his expected payoff is 
\begin{align*}
	E[u|x,d=1]=\frac{1}{2}(\gamma_G(Y-D_1)+X)+\frac{1}{2}(\gamma_B(Y-D_1)).
\end{align*}

When the manager withholds $x$, his payoff is 
\begin{align*}
	E[u|x,d=0]=&\;\frac{1}{2}(1-x)(\gamma_G(Y-D_0)+X)+\frac{1}{2}
	x(\gamma_G(Y-D_0)+\tau(1-\kappa)L_G) \\
	&+\frac{1}{2}\gamma_B(Y-D_{0}).
\end{align*}

Therefore, the incremental payoff of disclosing relative to withholding,
denoted by $\Delta(x)$, is 
\begin{align}  \label{eqn:incremental2}
	\Delta(x)=\frac{1}{2}(\gamma_G+\gamma_B)(D_0-D_1)+\frac{1}{2}x(X-\tau
	(1-\kappa )L_{G}).
\end{align}

Combining the two cases, we have 
\begin{align}  \label{eqn:incremental}
	\Delta(x)= 
	\begin{cases}
		\frac{1}{2}(\gamma_G+\gamma_B)(D_0-D_1)+\frac{1}{2}x(X+\tau (1-\kappa
		)L_{B}), \text{ if } x\le0 \\ 
		\frac{1}{2}(\gamma_G+\gamma_B)(D_0-D_1)+\frac{1}{2}x(X-\tau (1-\kappa
		)L_{G}), \text{ if } x>0.%
	\end{cases}%
\end{align}
In equilibrium, $D_0$ and $D_1$ only depend on the lender's conjecture of $x$
and therefore is independent of the realization of $x$. Thus, $\Delta(x)$ is
continuous and strictly increasing in $x$. Since the manager discloses $x$
if and only if $\Delta(x)>0$, the equilibrium must be a threshold strategy
where the manager discloses $x$ if and only if $x>x^*$.

\subsubsection{Step 2: Prove that \texorpdfstring{$x^*<0$}{Lg}}

We now calculate $D_0$ and $D_1$ that make the lender break-even. When $x$
is disclosed, GAAP will be replaced by a perfect rule $R^{\prime }$ with no
errors. As a result, $D_1$ must satisfy 
\begin{align}  \label{eqn:D1 break even}
	\frac{1}{2}\gamma_GD_1+\frac{1}{2}\gamma_BD_1+\frac{1}{2}(1-\gamma_B)y=K.
\end{align}
Solving the above equation gives 
\begin{align}  \label{eqn:D1}
	D_1=\frac{K-\frac{1}{2}(1-\gamma_B)y}{\frac{1}{2}\gamma_G+\frac{1}{2}%
		\gamma_B }.
\end{align}

When there is no disclosure, the lender rationally anticipates that the
manager has withheld $x$ or the manager failed to uncover $x$. Thus, the
lender's expected payoff conditional on non-disclosure is 
\begin{align}  \label{eqn:lender}
	E[v|m=\emptyset]=\frac{1-p}{1-p+p\int_{-1}^{x^*}f(x)dx}E[v|t=\emptyset]+ 
	\frac{p\int_{-1}^{x^*}f(x)dx}{1-p+p\int_{-1}^{x^*}f(x)dx}E[v|x\in[-1,x^*]].
\end{align}
The term $E[v|t=\emptyset]$ is the lender's average payoff across all false
alarm and undue optimism errors: 
\begin{align*}
	E[v|t=\emptyset]=&\int_{-1}^{0}(\frac{1}{2}(\gamma_{G}+\gamma_B)D_0+\frac{1}{
		2}(1-\gamma_B)y+\frac{1}{2}(-x)((1-\tau)(1-\kappa)L_B-(1-\gamma_B)y))f(x)dx
	\\
	+&\int_{0}^{1}(\frac{1}{2}(\gamma_{G}+\gamma_B)D_0+\frac{1}{2}(1-\gamma_B)y+ 
	\frac{1}{2}x((1-\gamma_G)y+(1-\tau)(1-\kappa)L_G))f(x)dx \\
	=&\frac{1}{2}(\gamma_{G}+\gamma_B)D_0+\frac{1}{2}(1-\gamma_B)y \\
	+&\int_{0}^{1}\frac{1}{2}((1-\tau)(1-\kappa)(\gamma_G-\gamma_B)y-(\gamma_G-
	\gamma_B)y)xf(x)dx,
\end{align*}
where the first equality applies payoffs from Tables \ref{Table:fg_t=2} and %
\ref{Table:fb_t=2}, and the second equality uses the assumption that $f(x)$
is an even function so $f(x)=f(-x)$ for any $x$. Since $(1-\tau)(1-\kappa)(%
\gamma_G-\gamma_B)y-(\gamma_G-\gamma_B)y<0$, we have 
\begin{align}  \label{eqn:evempty}
	E[v|t=\emptyset]<\frac{1}{2}(\gamma_{G}+\gamma_B)D_0+\frac{1}{2}
	(1-\gamma_B)y.
\end{align}
The above inequality further implies that 
\begin{align}  \label{eqn:evx}
	E[v|x\in[-1,x^*]]\le E[v|t=\emptyset]<\frac{1}{2}(\gamma_{G}+\gamma_B)D_0+ 
	\frac{1}{2}(1-\gamma_B)y.
\end{align}
The first inequality in \eqref{eqn:evx} holds because the lender's payoff is
increasing in $x$, and therefore, the truncated mean must be lower than the
unconditional mean.

Based on \eqref{eqn:lender}, \eqref{eqn:evempty}, and \eqref{eqn:evx}, we
have 
\begin{align}  \label{eqn:evm}
	E[v|m=\emptyset]<\frac{1}{2}(\gamma_{G}+\gamma_B)D_0+\frac{1}{2}
	(1-\gamma_B)y.
\end{align}
In addition, since the lender breaks-even, we have $E[v|m=\emptyset]=K$.
Thus, comparing \eqref{eqn:D1 break even} and \eqref{eqn:evm} lead to $%
D_0>D_1$.

At the threshold $x^*$, the manager must be indifferent between withholding
and disclosing. Suppose $x^*>0$, then the indifference condition would be 
\begin{align*}
	\Delta(x^*)=\frac{1}{2}(\gamma_G+\gamma_B)(D_0-D_1)+\frac{1}{2}x^*(X-\tau
	(1-\kappa )L_{G})=0.
\end{align*}
However, since $D_0>D_1$ and $X-\tau (1-\kappa )L_{G}>0$, the above equation
has no solution with $x^*>0$, a contradiction. Therefore, we must have $%
x^*<0 $ in equilibrium.

\subsection{Proof of Proposition \ref{prop:threshold}}

\subsubsection{Step 1: Simplify the indifference condition}

We know from Proposition \ref{prop:D} that $x^*<0$. Thus, the indifference
condition is 
\begin{align}  \label{eqn:indiff}
	\Delta(x^*)=\frac{1}{2}(\gamma_G+\gamma_B)(D_0-D_1)+\frac{1}{2}x^*(X+\tau
	(1-\kappa )L_{B})=0.
\end{align}
Recall that $D_1$ is determined by equation \eqref{eqn:D1 break even}. $D_0$
is determined by the following break-even condition 
\begin{align}  \label{eqn:D0 breakeven}
	E[v|m=\emptyset]=\frac{1-p}{1-p+p\int_{-1}^{x^*}f(x)dx}E[v|t=\emptyset]+ 
	\frac{p\int_{-1}^{x^*}f(x)dx}{1-p+p\int_{-1}^{x^*}f(x)dx}E[v|x\in[-1,x^*]]=K.
\end{align}
We know from the proof of Proposition \ref{prop:D} that 
\begin{align*}
	E[v|t=\emptyset]=\frac{1}{2}(\gamma_{G}+\gamma_B)D_0+\frac{1}{2}
	(1-\gamma_B)y+\int_{0}^{1}\frac{1}{2}((1-\tau)(1-\kappa)(\gamma_G-
	\gamma_B)y-(\gamma_G-\gamma_B)y)xf(x)dx.
\end{align*}
In addition, 
\begin{align*}
	&E[v|x\in[-1,x^*]]= \\
	&\int_{-1}^{x^*}(\frac{1}{2}(\gamma_{G}+\gamma_B)D_0+\frac{1}{2}
	(1-\gamma_B)y+\frac{1}{2}(-x)((1-\tau)(1-\kappa)L_B-(1-\gamma_B)y))\frac{%
		f(x) }{\int_{-1}^{x^*}f(x)}dx.
\end{align*}
Based on the above expressions, we can simplify \eqref{eqn:D0 breakeven} to 
\begin{align}
	&\frac{1}{2}(\gamma_{G}+\gamma_B)D_0+\frac{1}{2}(1-\gamma_B)y+  \notag \\
	&\frac{p}{1-p+p\int_{-1}^{x^*}f(x)dx}\int_{-1}^{x^*}\frac{1}{2}
	(-x)((1-\tau)(1-\kappa)L_B-(1-\gamma_B)y)f(x)dx+  \notag \\
	&\frac{1-p}{1-p+p\int_{-1}^{x^*}f(x)dx}C=K,  \label{eqn:D0}
\end{align}
where $C$ is a negative constant that does not depend on $x^*$: 
\begin{align}  \label{eqn:C}
	C=\int_{0}^{1}\frac{1}{2}((1-\tau)(1-\kappa)(\gamma_G-\gamma_B)y-(\gamma_G-
	\gamma_B)y)xf(x)dx<0.
\end{align}
Subtracting \eqref{eqn:D1 break even} from \eqref{eqn:D0} lead to 
\begin{align}
	&\frac{1}{2}(\gamma_{G}+\gamma_B)(D_0-D_1)=  \notag \\
	&\frac{p}{1-p+p\int_{-1}^{x^*}f(x)dx}\int_{-1}^{x^*}\frac{1}{2}
	x((1-\tau)(1-\kappa)L_B-(1-\gamma_B)y)f(x)dx-  \notag \\
	&\frac{1-p}{1-p+p\int_{-1}^{x^*}f(x)dx}C.  \label{eqn:diff}
\end{align}
Plugging \eqref{eqn:diff} into the indifference condition \eqref{eqn:indiff}
leads to 
\begin{align*}
	&\frac{p}{1-p+p\int_{-1}^{x^*}f(x)dx}\int_{-1}^{x^*}\frac{1}{2}
	x((1-\gamma_B)y-(1-\tau)(1-\kappa)L_B)f(x)dx+ \\
	&\frac{1-p}{1-p+p\int_{-1}^{x^*}f(x)dx}C=\frac{1}{2}x^*(X+\tau (1-\kappa
	)L_{B}).
\end{align*}
Defining 
\begin{align}  \label{eqn:C1}
	C_1=\frac{1}{2}((1-\gamma_B)y-(1-\tau)(1-\kappa)L_B)>0
\end{align}
and 
\begin{align}  \label{eqn:C2}
	C_2=\frac{1}{2}(X+\tau (1-\kappa )L_{B})>0,
\end{align}
and the indifference condition simplifies to 
\begin{align}  \label{eqn:indiffsimplified}
	\frac{p}{1-p+p\int_{-1}^{x^*}f(x)dx}\int_{-1}^{x^*}C_1xf(x)dx+\frac{1-p}{
		1-p+p\int_{-1}^{x^*}f(x)dx}C=C_2x^*.
\end{align}
Using integration by parts, we can transform \eqref{eqn:indiffsimplified} to 
\begin{align}  \label{eqn:indifffinal}
	pC_1\int_{-1}^{x^*}F(x)dx+p(C_2-C_1)x^*F(x^*)+(1-p)(C_2x^*-C)=0.
\end{align}

\subsubsection{Step 2: The existence and uniqueness of the equilibrium}

Denote the right-hand side of \eqref{eqn:indifffinal} as $J(x^*)$. Note that 
$C_2-C_1=-\frac{1}{2}\kappa L_B$. Thus, when $\kappa$ is sufficiently small, 
$J^{\prime }(x^*)=pC_1F(x^*)+(1-p)C_2+p(C_2-C_1)(F(x^*)+x^*f(x^*))>0$, i.e., 
$J(x^*)$ strictly increases in $x^*$.

In addition, we have $J(x^*=0)=pC_1\int_{-1}^{0}F(x)dx-(1-p)C>0$, as $C<0$.
We also have $J(x^*=-1)=(1-p)(-C_2-C)$. Thus $J(x^*=-1)$ has the opposite
sign to $C_2+C$. Based on the expressions of $C$ and $C_2$ in \eqref{eqn:C}
and \eqref{eqn:C2}, $C_2+C>0$ if and only if 
\begin{align}  \label{eqn:comparison}
	X+\tau (1-\kappa
	)L_{B}>(1-(1-\tau)(1-\kappa))(\gamma_G-\gamma_B)y\int_{0}^{1}xf(x)dx.
\end{align}

When $\kappa=0$, the left hand side of \eqref{eqn:comparison} is equal to $%
(1-\tau)X+\tau(1-\gamma_B)y$, whereas the right hand side of %
\eqref{eqn:comparison} is equal to $\tau(\gamma_G-\gamma_B)y%
\int_{0}^{1}xf(x) $.

Note that $E[x|x\in[0,1]]=\frac{\int_0^1xf(x)dx}{F(1)-F(0)}<1$, which
implies that $\int_0^1xf(x)dx<F(1)-F(0)<1$. Comparing the left hand side of %
\eqref{eqn:comparison} with the right hand side of \eqref{eqn:comparison}
leads to 
\begin{align*}
	(1-\tau)X+\tau(1-\gamma_B)y>\tau(\gamma_G-\gamma_B)y\int_{0}^{1}xf(x),
\end{align*}
as $\gamma_G<1$. By continuity, $C_2+C>0$ (i.e., $J(x^*=-1)<0$) when $\kappa$
is sufficiently small. By the Intermediate Value Theorem, we know that %
\eqref{eqn:indifffinal} has a unique solution $x^*\in[-1,0]$ when $\kappa$
is sufficiently small.

\subsection{Proof of Proposition \ref{prop:cs}}

We know from \eqref{CSalterna} from the main text that 
\begin{align*}
	\frac{\partial x^*}{\partial \beta}&= \frac{\frac{\partial (1-p)C}{\partial
			\beta}-\frac{\partial pC_1}{\partial \beta}\int_{-1}^{x^*}F(x)dx-x^*\frac{
			\partial(1-p)C_2}{\partial \beta}}{pC_1F(x^*)+(1-p)C_2} \\
	&\propto \frac{\partial (1-p)C}{\partial \beta}-\frac{\partial pC_1}{
		\partial\beta}\int_{-1}^{x^*}F(x)dx-x^*\frac{\partial(1-p)C_2}{\partial
		\beta }.
\end{align*}

Since $x^*<0$ and based on Table \ref{Table: CS1}, when $\frac{\partial C_1}{
	\partial \beta}$ and $\frac{\partial C_2}{\partial \beta}$ have the same
sign, the sign of $\frac{\partial x^*}{\partial \beta}$ is ambiguous. Thus,
the prediction is ambiguous for $\gamma_B$, $y$, $X$, and $\tau$. It is
straightforward that $\frac{\partial x^*}{\partial p}<0$ and $\frac{\partial x^*}{
	\partial \kappa}<0$.

When $x$ follows a uniform distribution on $[-1,1]$, we have $f(x)=\frac{1}{2%
}$, $F(x)=\frac{1}{2}x+1/2$, and $\int_{-1}^{x^*}F(x)=\frac{1}{4}(1+x^*)^2$.

Equation \eqref{eqn:indiffmain} becomes: 
\begin{align*}
	\frac{1}{4}p C_1 (1 + x^*)^2 + p(C_2 - C_1)x^*(\frac{1}{2}x^*+ \frac{1}{2})
	+ (1-p)(C_2x^* - C)=0.
\end{align*}
Solving the equation gives 
\begin{align*}
	x^*=-\frac{C_2 p+\sqrt{C_2^2 (p-2)^2+\left(C_1-2 C_2\right) p \left(4 C
			(p-1)+C_1 p\right)}-2 C_2}{\left(C_1-2 C_2\right) p}.
\end{align*}
Computing the derivatives and letting $\kappa\rightarrow 0$ leads to 
\begin{align*}
	&\lim_{\kappa \rightarrow 0} \frac{\partial x^*}{\partial \tau}=\frac{1}{
		2\left(\tau y \left(\gamma _B-1\right)+(\tau -1) X\right) }\times \\
	&\frac{(p-1) X y \left(\gamma _B-\gamma _G\right)}{\sqrt{(p-1) \left(\tau y
			\left(\gamma _B-1\right)+(\tau -1) X\right) \left(\tau y \left((p-2) \gamma
			_B-p \gamma _G+2\right)-2 (\tau -1) X\right)}}<0. \\
	&\lim_{\kappa \rightarrow 0} \frac{\partial x^*}{\partial \gamma_B}=\frac{1}{
		2\left(\tau y \left(\gamma _B-1\right)+(\tau -1) X\right) }\times \\
	&-\frac{(p-1) \tau y \left(\tau y \left(\gamma _G-1\right)+(\tau -1)
		X\right) }{\sqrt{(p-1) \left(\tau y \left(\gamma _B-1\right)+(\tau -1)
			X\right) \left(\tau y \left((p-2) \gamma _B-p \gamma _G+2\right)-2 (\tau -1)
			X\right)} }>0. \\
	&\lim_{\kappa \rightarrow 0} \frac{\partial x^*}{\partial X}=\frac{1}{
		2\left(\tau y \left(\gamma _B-1\right)+(\tau -1) X\right) }\times \\
	&\frac{(p-1) (\tau -1) \tau y \left(\gamma _B-\gamma _G\right)}{\sqrt{(p-1)
			\left(\tau y \left(\gamma _B-1\right)+(\tau -1) X\right) \left(\tau y
			\left((p-2) \gamma _B-p \gamma _G+2\right)-2 (\tau -1) X\right)}}>0. \\
	&\lim_{\kappa \rightarrow 0} \frac{\partial x^*}{\partial y}=\frac{1}{
		2\left(\tau y \left(\gamma _B-1\right)+(\tau -1) X\right) }\times \\
	&-\frac{(p-1) (\tau -1) \tau X \left(\gamma _B-\gamma _G\right)}{\sqrt{(p-1)
			\left(\tau y \left(\gamma _B-1\right)+(\tau -1) X\right) \left(\tau y
			\left((p-2) \gamma _B-p \gamma _G+2\right)-2 (\tau -1) X\right)}}<0.
\end{align*}
By continuity, all the above comparative statics hold for sufficiently small 
$\kappa$.

\subsection{Proof of Proposition \protect\ref{prop:effort}}

We first determine the equilibrium effort $p^*$. When the manager is
uninformed, his expected payoff is equal to first-best total surplus minus
the cost of renegotiation minus the lender's payoff. 
\begin{align}  \label{eqn:uninformed}
	E[u|t=\emptyset]=W_{FB}-\frac{1}{2}\int_0^1(\kappa L_G+\kappa
	L_B)xf(x)dx-E[v|t=\emptyset].
\end{align}

When the manager is informed and $x>x^*$, his expected payoff is 
\begin{align}  \label{eqn:disclose}
	E[u|t=x,m=x]=W_{FB}-K.
\end{align}
as the lender breaks even and receives $K$.

When the manager is informed and $x<x^*$, his expected payoff satisfies 
\begin{align}  \label{eqn:withhold}
	E[u|t=x,m=\emptyset]>W_{FB}-K,
\end{align}
as the manager must obtain a higher payoff by withholding $x$.

By exerting effort $p$, the manager's expected payoff is: 
\begin{align*}
	E[u]=(1-p)E[u|t=\emptyset]+p\int_{-1}^{x^*}E[u|t=x,m=\emptyset]f(x)dx+p
	\int_{x^*}^{1}E[u|t=x,m=x] f(x)dx.
\end{align*}
The first order condition is 
\begin{align}
	c'(p^*)&=\int_{-1}^{x^*}E[u|t=x,m=\emptyset]f(x)dx+
	\int_{x^*}^{1}E[u|t=x,m=x]f(x)dx-E[u|t=\emptyset]  \notag\\
	&>W_{FB}-K-(W_{FB}-\frac{1}{2}\int_0^1(\kappa L_G+\kappa
	L_B)xf(x)dx-E[v|t=\emptyset])  \notag \\
	&=\frac{1}{2}\int_0^1(\kappa L_G+\kappa L_B)xf(x)dx+E[v|t=\emptyset]-K 
	\notag \\
	&>\frac{1}{2}\int_0^1(\kappa L_G+\kappa L_B)xf(x)dx  \notag \\
	&=c'(p^{FB}).  \notag
\end{align}

The first inequality utilizes \eqref{eqn:uninformed}-\eqref{eqn:withhold}.
Based on equation \eqref{eqn:D0 breakeven} and the observation that lender
breaks even (i.e., $E[v|m=\emptyset]=K$), we must have $E[v|t=\emptyset]>K$
and thus the second inequality holds. The last line holds due to equation %
\eqref{FOC}. Since $c(.)$ is convex and $c'(p^*)>c'(p^{FB})$, we
know that $p^*>p^{FB}$.

\bibliographystyle{chicago}
\bibliography{bib}

\end{document}